\documentclass[amsfonts]{article}
\usepackage{graphicx}
\pdfoutput=1
\usepackage{subcaption}

\newcommand{\be}{\begin{equation}}
\newcommand{\ee}{\end{equation}}
\newcommand{\bea}{\begin{array}}
\newcommand{\ea}{\end{array}}
\newcommand{\beqa}{\begin{eqnarray}}
\newcommand{\eeqa}{\end{eqnarray}}

\newcommand{\eean}{\end{eqnarray*}}

\def\up#1{\leavevmode \raise.16ex\hbox{#1}}

\newcommand{\gapproxeq}{\lower
 .7ex\hbox{$\;\stackrel{\textstyle >}{\sim}\;$}}
\newcommand{\lapproxeq}{\lower .7ex\hbox{$\;\stackrel
{\textstyle <}{\sim}\;$}}

\newcounter{appendice}

\def\thebibliography#1{{\bf REFERENCES\markboth
 {REFERENCES}{REFERENCES}}\list
 {[\arabic{enumi}]}{\settowidth\labelwidth{[#1]}\leftmargin\labelwidth
 \advance\leftmargin\labelsep
 \usecounter{enumi}}
 \def\newblock{\hskip .11em plus .33em minus -.07em}
 \sloppy
 \sfcode`\.=1000\relax}

\def\BI{{\rm 1\!l}}

\begin{document}

\centerline{ \LARGE Matrix Model Cosmology in Two Space-time Dimensions}

\vskip 2cm

\centerline{ A. Stern\footnote{astern@ua.edu}   }

\vskip 1cm
\begin{center}
  { Department of Physics, University of Alabama,\\ Tuscaloosa,
Alabama 35487, USA\\}

\end{center}
\vskip 2cm

\vspace*{5mm} 

\normalsize
\centerline{\bf ABSTRACT}

\noindent
We  examine   solutions to the classical IKKT matrix model equations in three space-time dimensions. Closed, open and static two-dimensional universes naturally emerge from such models in the commutative limit. We show that tachyonic modes are a generic feature of  these cosmological solutions.

\newpage

\section{Introduction}
\setcounter{equation}{0}
Matrix models promise   to be a convenient tool for studying  nonperturbative aspects of string theory.\cite{Ishibashi:1996xs},\cite{Banks:1996vh}                                                                
  Space-time geometry,   field theory and  gravity can dynamically  emerge from such models,\cite{Aoki:1998vn},\cite{Steinacker:2010rh} and thus  can have implications in  cosmology.\cite{Freedman:2004xg},\cite{Craps:2005wd},\cite{Klammer:2009ku},\cite{Kim:2011ts}.  The matrix model approach to cosmology has the advantage of including nonperturbative string theory effects and possibly resolving cosmological singularities.\cite{Craps:2005wd}  Previously, numerical simulations were used to show  how a three-spatial dimensional expanding universe can emerge from a ten dimensional matrix model.\cite{Kim:2011ts}
Moreover, solutions  to the classical equations of motion have been found which resemble expanding universes,\cite{Freedman:2004xg},\cite{Kim:2011ts} and can support many desirable features, such as  a big bounce and an early inflationary phase with graceful
exit.\cite{Klammer:2009ku}

While the appropriate  framework for matrix models is a  ten dimensional supersymmetric theory (either IKKT\cite{Ishibashi:1996xs} or BFSS\cite{Banks:1996vh}),  an examination of simpler systems may prove beneficial.   With this in mind, we shall  restrict our attention to the bosonic sector of  the IKKT  matrix model in three space-time dimensions.  We write down  the standard matrix equations in section 2, and focus on its classical solutions. Evidence for  nontrivial solutions to the matrix model equations is seen  by going  to  the commutative limit, where  the classical equations of motion coincide with those of a closed Nambu string, and are easily solved.  One is the cylindrically symmetric solution, which corresponds to a closed two-dimensional  surface with  an initial and final singularity.  (Such singularities are not expected to appear in the analogous matrix model solution.) This  solution  is the Lorentzian space counterpart to the minimum area  catenoid in Euclidean space, whose corresponding matrix model solution is nontrivial  because it does not correspond to a finite dimensional Lie algebra.\cite{Arnlind:2012cx}  Similarly, the matrix model solution in the  Minkowski  space background (assuming it exists) which gives the cylindrically symmetric  closed two-dimensional  surface  in the commutative limit, is nontrivial for the same reason.  However, in this article  our aim is not in finding an expression for the matrix   solution.  We shall instead be  examining stability questions, more specifically, in the commutative limit.  Perturbations about the solution can be expressed in terms of an abelian gauge field and scalar field (or nonabelian gauge fields and $N$ scalar fields if one expands about a stack of $N$ coinciding branes).  This is possible thanks to the  use of a Seiberg-Witten map\cite{Seiberg:1999vs} on the noncommutative space associated with the solution.  We obtain the map up to first order in the noncommutativity parameter in order to obtain the lowest order efffects in the action. 
Gauge transformations correspond to area preserving coordinate transformations on the two-dimensional surface, while the scalar field is associated with perturbations normal to the surface.   At leading order, the perturbed action yields the usual description of a scalar field, which is decoupled to the (nondynamical) gauge field.
We find that the scalar field is tachyonic, and  thus that the system is unstable with respect to perturbations normal to the surface.

The system can be generalized with the inclusion of  cubic term in the matrix model action, and we do this in section 3. This term is the matrix analogue of a topological term.  It preserves the symmetries of the three-dimensional matrix model and introduces  a free parameter $\upsilon$ in the theory. Three different types of nontrivial cylindrically symmetric  solutions of the matrix equations can result from this model.    
Two are well known, and they are associated with finite dimensional Lie algebras.  One  is  the noncommutative   de Sitter solution with the associated algebra beings $so(2,1)$.\cite{Ho:2000fy},\cite{Jurman:2013ota}.  This solution   is the Lorentzian space analogue of the fuzzy sphere.\cite{Madore:1991bw}-\cite{Iso:2001mg} 
Another solution is the noncommutative cylinder, whose associated algebra generates the two dimensional Euclidean group.\cite{Chaichian:2000ia},\cite{Balachandran:2004yh},\cite{me}  Neither of these two solutions are present when the cubic term is removed, corresponding to the $\upsilon\rightarrow 0$ limit.   Evidence for the existence of a third class of  solutions can be seen  by  once again going  to  the commutative limit.   In that limit, we obtain solutions which are a deformation of the cylindrically symmetric solution to the Nambu string described above.   Like with the previous string solution, they are not  associated with any finite dimensional Lie algebra, and so their matrix model analogues are nontrivial.  Here one gets a continuous family of cylindrically symmetric solutions  (parametrized by $\upsilon$) which can describe closed, stationary or open space-times, the choice depending on the value of $\upsilon$.  A string energy-momentum tensor can be defined  for this system, and the energies of all the solutions can be compared.  In this regard, we find that the solutions corresponding to open and stationary space-times are energetically favored.  We also perform  perturbations about the different solution, and  again  express  them in terms of an abelian gauge field and scalar field.  At leading order, the action now reveals a coupling between the gauge field and scalar, which is not present when $\upsilon=0$.  We also find that the effective mass-squared for the scalar field is negative for all of the  solutions (and moreover, it can be scale-dependent).  Thus as before, these systems are unstable with respect to perturbations  normal to the surface.  

Possible generalizations and cures of the instabilities are discussed in section 4.

Note added:  The work presented here is similar in spirit to that of Klammer and Steinacker\cite{Klammer:2009ku}, where four-dimensional cosmological solutions were found to have many desirable features.
 Our work differs in that we do not need to construct alternatives to the standard induced  metric in two space-time dimensions and we require no Wick rotations. The instabilities which we find here are independent of the choice of metric tensor.

\section{Lorentzian matrix model}
\setcounter{equation}{0}
\subsection{Classical equations and the commutative limit}

We consider the IKKT Lorentzian matrix model associated with three space-time dimensions.  As here we shall only be concerned with the bosonic sector,
the dynamical degrees of freedom  are contained in three infinite-dimensional Hermitean matrices, which we denote by  $Y^\mu$, $\mu=0,1,2$.   For the action $S(Y)$,  we have  the usual quartic term
\be S(Y)=-\frac 1{4g^2}{\rm Tr}\; [Y_\mu, Y_\nu] [Y^\mu,Y^\nu] \;,\label{mmactnnqt}\ee  $g$ being a constant,
with resulting equations of motion
\be [[ Y_\mu,Y_\nu],Y^\nu] =0\label{eqofmotnqt} \ee
 We raise and lower indices with the flat metric $\eta_{\mu\nu}={\rm diag}(- 1,1,1)$. 
 The equations of motion (\ref{eqofmotnqt}) are  invariant under: i) 
 Lorentz transformations $Y^\mu\rightarrow L^\mu_{\;\;\nu} Y^\nu$,  where $L$ is a $3\times 3$ Lorentz matrix,
 ii)  translations in the three-dimensional Minkowski space $Y^\mu\rightarrow Y^\mu+v^\mu\BI$, where $\BI$ is the  unit matrix, and  iii) unitary `gauge' transformations, $Y^\mu\rightarrow UY^\mu U^\dagger$, where $U$ is an infinite dimensional unitary matrix. 

Evidence for a nontrivial solution to (\ref{eqofmotnqt})is seen by going to the commutative limit of the matrix model.  
The commutative limit corresponds to the replacement of  the matrices $Y^\mu$, $\mu=0,1,2$, by  space-time coordinates $y^\mu$, $\mu=0,1,2$, and the replacement of  the commutator of  functions of  $Y^\mu$  with some Poisson bracket of functions of $y^\mu$.  For this we can introduce a noncommutativity parameter $\theta$, with    the commutative limit corresponding to  $\theta\rightarrow 0 $. To lowest order in   $\theta$,
$[f(Y),h(Y)]\rightarrow i\theta \{f(y),h(y)\}$, $\{\;,\;\}$ denoting the Poisson bracket which is required to satisfy the usual properties such as the Jacobi identity.  In the commutative limit,  the equations of motion  (\ref{eqofmotnqt})  take the form
\be \{\{y_\mu,y_\nu\},y^\nu\} =0 \label{clmeasnqt}
\;\ee  

\subsection{A classical string solution}  

Thruought this article we shall be considering    solutions  corresponding to  cylindrically symmetric  surfaces, with $x^0$  (time) as the central axis.  If $y^\mu= x^\mu$ denotes such a solution, we can  write 
\be  ( x^1)^2+ ( x^2)^2=a^2(x^0)\;\label{cmtvcnstrnt}\ee
The function $a$ is the radius of any time-slice and it plays the role of the scale factor.
In addition to  $a$, we introduce  the function  $h$ of $x^0$ in the  following expression for the  Poisson brackets of the coordinates:
\beqa  \{x^1,x^2\}&=&h(x^0)\,a(x^0)a'(x^0)\cr
 \{x^2,x^0\}&=&-h(x^0)\,x^1\cr
 \{x^0,x^1\}&=&-h(x^0)\,x^2\label{gnrlpbs}\;,\eeqa
where the prime denotes differentiation with respect to $x^0$.
The Poisson brackets are consistent with the constraint (\ref{cmtvcnstrnt}) and 
identically satisfy the Jacobi identity.  They solve the equations of motion  (\ref{clmeasnqt}) provided that the two functions  $a$ and $h$ satisfy
\be  \Bigl( (aa'h)'+h\Bigr)\,h=0\qquad \quad \Bigl(2a'h+ah'\Bigr)\,ah=0\label{3difeqsnqt}\ee
 For $h\ne0$, we get the following equation for the scale factor
 \be \frac {a''} a=\Bigl(\frac {a'}a\Big)^2-\frac 1{a^2}\ee The integral of motion is  $a/\sqrt{1-{a'}^2}$, which  we shall see later is associated with the energy of a bosonic string.  The equations (\ref{3difeqsnqt}) are  easily solved by
 \be a(x^0)=\cos x^0\qquad\quad h(x^0)=\sec^2 x^0\;,\label{afhnqt}\ee where
$\;- \frac \pi 2\le x^0\le\frac \pi 2$.   They are consistent with the boundary values $a(0)=1$ and $a'(0)=0$. The corresponding surface is a closed two-dimensional space-time with an initial and final singularity at $ x^0=-\frac \pi 2$ and  $ x^0=\frac \pi 2$, respectively.  Singularities are not expected to appear in the corresponding matrix solution.  The surface is pictured in figure 1(b).

The above solution  is the Lorentzian space counterpart to the minimum area  catenoid in Euclidean space.
The matrix analogue of the catenoid in Euclidean space is nontrivial,  as it is not associated with any finite dimensional Lie algebra.\cite{Arnlind:2012cx}  The same holds for the matrix solution in the Minkowski space background, and methods similar to those used in \cite{Arnlind:2012cx}  can be applied to obtain it.  Our  interest here is to instead examine stability questions about the solution.  For this we restrict our attention to the commutative limit.

In order to examine stability, consider a family of two-dimensional closed  surfaces  $y^\mu=y_\epsilon^\mu(\tau,\sigma)$ embedded in three-dimensional Minkowski space, where $\epsilon$ parametrizing the different surfaces, while  $\tau$ and $\sigma$, $0\le \sigma<2 \pi$, are the time and space parameters,  respectively,  spanning any given surface. 
Let the surface with $\epsilon=0$ correspond to  the classical solution  (\ref{afhnqt}), $y^\mu_0(\tau,\sigma)=x^\mu(\tau,\sigma)$.
 An explicit parametrization of the solution $x^\mu(\tau,\sigma)$ is
\be \pmatrix{x^0\cr x^1\cr x^2}= \pmatrix{\tau\cr a( \tau)\cos\sigma\cr a( \tau)\sin\sigma}\;,\label{ncprtztn}\ee which only holds for the restricted time domain $ - \frac \pi 2\le \tau\le\frac \pi 2$. We recover 
 (\ref{gnrlpbs}) upon defining the Poisson brackets of any two functions ${\cal F}$ and ${\cal G}$ of $\tau$ and $ e^{i\sigma}$ according to
\be  \{{\cal F},{\cal G} \}(\tau, e^{i\sigma})= h(\tau)\Bigl(\partial_\tau {\cal F} \partial_\sigma {\cal G} - \partial_\sigma{\cal F} \partial_\tau {\cal G}\Bigr)\label{pbtauphi}\ee  More generally, we can  use  (\ref{pbtauphi}), with $h$ replaced by a general function of both parameters, to define the Poisson brackets on any of the closed two-dimensional surfaces $y^\mu=y_\epsilon^\mu(\tau,\sigma)$.

In addition to being a solution of (\ref{clmeasnqt}), 
 (\ref{ncprtztn}) also solves the equations of motion for a classical closed bosonic string.  More generally,  (\ref{clmeasnqt}) contain the string equations of motion.  For this one can introduce 
 the induced metric \be {\tt g}_{{\tt ab}}(\tau,\sigma)=\partial_{\tt a}y^\mu\partial_{\tt b} y_\mu\;,\qquad\;{\tt a}=\tau,\sigma\;,\label{ndcdmtrc}\ee
on the two-dimensional surface or world sheet.  The standard  Nambu-Goto action is
\be S_{NG}=-{\cal T}\int d\tau d\sigma\, \sqrt{-{\tt g}}\;,\label{clstactn}\ee where ${\tt g}$ is the determinant of the induced metric, and the constant ${\cal T}$ denotes the string tension. 
The equations of motion resulting from (\ref{clstactn}) are 
\be  \Delta y_\mu =0\;,\label{cleomnqt}\ee where $\Delta =-\frac 1{\sqrt{-{\tt g}}}\partial_{\tt a}\sqrt{-{\tt g}} {\tt g}^{\tt ab}\partial_{\tt b}$ is the Laplace-Beltrami operator on the world sheet, ${\tt g}^{\tt ab}$ denotes the components of the inverse induced metric, ${\tt g}^{\tt ab}{\tt  g}_{\tt bc}=\delta ^{\tt a}_{\tt c}$.   The string equations (\ref{cleomnqt}) are identical to (\ref{clmeasnqt}) when  the Poisson structure on the world sheet involves the metric tensor, specifically\cite{Arnlind:2012cx} 
\be  \{{\cal F},{\cal G} \}(\tau, e^{i\sigma})=\frac 1{\sqrt{-{\tt g}}}\Bigl(\partial_\tau {\cal F} \partial_\sigma {\cal G} - \partial_\sigma{\cal F} \partial_\tau {\cal G}\Bigr)\;,\label{rltpbmtrc} \ee which leads to $\{y^\mu,y^\nu\}\{y_\mu,y_\nu\}=-2$.  So in comparing with (\ref{pbtauphi}), we get the condition 
\be h =\frac 1{\sqrt{-{\tt g}}} \label{his1vrsmt}\;,\ee
which is in fact satisfied for the solution  (\ref{afhnqt}).  

 The equations of motion (\ref{cleomnqt}) imply the existence of a conserved current $p^{\tt a}_\mu$ on the world sheet, $\partial_{\tt a}p^{\tt a}_\mu=0$, where
\be p^{\tt a}_\mu=-{\cal T}\sqrt{-{\tt g}} {\tt g}^{\tt ab}\partial_{\tt b }y_\mu\label{panu}
\ee  From $p^{\tt a}_\mu$ one can construct the  stress-energy tensor in the three-dimensional  embedding space, \be T^{\mu\nu}(z)=\int d\sigma  d\tau \, p^{ {\tt a}\nu}\partial_{\tt a}y^\mu \,\delta^3(z-y(\sigma,\tau))\;,\label{strsnrg}\ee satisfying $\frac\partial{\partial z^\mu} T^{\mu\nu}(z)=0$, and thus define an energy $E$ of the string at any given time $z^3$, 
\be E=\int dz^1dz^2\, T^{00}(z) \label{strngnrg}\ee
Upon evaluating (\ref{strngnrg}) for a solution  of the form  (\ref{ncprtztn}), one gets 
\be E=2\pi {\cal T}\frac {a(z^0)}{\sqrt{1-a'(z^0)^2}}\;, \ee
which using  (\ref{afhnqt}) gives $2\pi {\cal T}$. Compared to the vacuum,  the solution  (\ref{afhnqt}) is energetically disfavored (assuming ${\cal T}$ to be positive).

Alternatively, we can  address the issue  of  stability from the perspective of the matrix model (or at least, its commutative limit), which we do next. 

\subsection{Stability analysis using the Seiberg-Witten map}

Here we consider small perturbations about the above classical solution.  For this we shall utilize the commutative limit of the  matrix model action (\ref{mmactnnqt}).  It  is
\be
 S_c(y)=\frac 1{4g_c^2}\int d\mu(\tau,\sigma) \{y_\mu, y_\nu\}\{y^\mu,y^\nu\} \;,\label{cmtvlmtscnqt}
\ee
 where $g_c$ is the limiting value of $g$ and $ d\mu(\tau,\sigma)$ is an invariant  integration measure on the two-dimensional surface.  The latter is defined such that $\int d\mu(\tau,\sigma)\,\{{\cal F},{\cal G}\}\,{\cal H}$  $=\int d\mu(\tau,\sigma)\,{\cal F}\,\{{\cal G},{\cal H}\}$, for arbitrary functions ${\cal F},\,{\cal G}$ and ${\cal H}$ on the world sheet. 
So upon assuming Poisson brackets (\ref{pbtauphi}), we can use the measure $d\mu(\tau,\sigma)=d\tau d\sigma/h(\tau)$. Along with Lorentz and   translational invariance, the action is invariant under the commutative analogue of the unitary `gauge' transformations.  Here  infinitesimal gauge variations have the form $\delta y^\mu=\theta\{\Lambda,y^\mu\}$, where  $\theta$ again denotes the noncommutativity parameter and $\Lambda$ is an infinitesimal function on the world sheet.  The equations (\ref{clmeasnqt}) follow from extremizing $ S_c$ with respect to variations of $y^\mu$. Here we do not have to require the condition   (\ref{his1vrsmt}) to hold for arbitrary configurations $y^\mu(\tau,\sigma)$, and $\{y^\mu,y^\nu\}\{y_\mu,y_\nu\}=-2$ is only valid for the solutions to the equations of motion.

Next we wish to evaluate the action (\ref{cmtvlmtscnqt}) for small perturbations about the solution $x^\mu(\tau,\sigma)$, defined by (\ref{cmtvcnstrnt}) and (\ref{gnrlpbs}).  For this we set 
\be y^\mu=x^\mu+ \theta A^\mu \;\label{yxbta}\ee
where  the noncommutativity parameter $\theta$ can be also be regarded as a  perturbation parameter and $A^\mu$ are three functions on the world sheet. ($A^\mu$ are replaced by $3N$  fields   if one instead expands about a stack of $N$ coinciding branes.)
 The perturbations (\ref{yxbta})  induce nonvanishing  fluctuations in the induced metric tensor $ {\tt g}_{\tt ab}$ at first order in $\theta$, and thus  $A_\mu$ affect the space-time geometry.  These functions transform as noncommutative gauge  potentials up to first order in $\theta$.  Infinitesimal gauge variations of $A_\mu$ are given by
\be \delta A_\mu=\{\Lambda, x_\mu\} +\theta\{\Lambda, A_\mu\}\label{gvrtnsAmu}\ee
Using  the Poisson brackets (\ref{pbtauphi}),  gauge variations at zeroth order in $\theta$ are along the tangential directions of the surface,
$\delta A_\mu= h(\tau)\Bigl(\partial_\tau\Lambda\partial_\sigma x_\mu-\partial_\sigma\Lambda\partial_\tau x_\mu\Bigr)+{\cal O}(\theta)$.  These leading order gauge variations are equivalent to area preserving\footnote{It can be checked that the determinant of the induced metric is gauge invariant up to first order in $\theta$.} infinitesimal reparametrizations of the surface  $y(\tau,\sigma)$: \be  (\tau,\sigma)\rightarrow\Bigl(\tau-\theta h(\tau)\partial_\sigma\Lambda\,,\,\sigma+\theta h(\tau)\partial_\tau\Lambda\Bigr)\ee
If we now include  first order terms, and use the parametrization (\ref{ncprtztn}), the gauge variations can be written as
\beqa
 \delta A^0&=& - h(\tau)\partial_\sigma\Lambda +\theta h(\tau)\Bigl(\partial_\tau\Lambda\partial_\sigma A^0-\partial_\sigma\Lambda\partial_\tau A^0\Bigr)
\cr &&\cr 
 \delta A_\pm &=&h(\tau) \Bigl(\pm ia(\tau)\partial_\tau\Lambda - a'(\tau)\partial_\sigma\Lambda\Bigr)e^{\pm i\sigma} +\theta h(\tau)\Bigl(\partial_\tau\Lambda\partial_\sigma A_\pm-\partial_\sigma\Lambda\partial_\tau A_\pm\Bigr)\;,\label{oneonesix}
\eeqa
where $A_\pm = A_1\pm A_2$.

  Using a Seiberg-Witten map\cite{Seiberg:1999vs}, the noncommutative potentials $A_\mu$ can be re-expressed in terms of commutative gauge potentials, denoted by  $({\cal A}_\tau,{\cal A}_\sigma)$, on  the surface, along with their derivatives.   Known expressions for the Seiberg-Witten map
on the Moyal plane\cite{swmaps}  do not apply in this case since the map must be consistent with the Poisson bracket relations (\ref{gnrlpbs}).  Moreover, since  the noncommutative potentials $A_\mu$ have three components  and  the commutative potentials have only two an additional  degree of freedom, associated with a scalar field $\phi$ should be included in the map.  Thus $A_\mu=A_\mu[{\cal A}_\tau,{\cal A}_\sigma,\phi]$.  Using  the  Seiberg-Witten map, commutative gauge transformation, $({\cal A}_\tau,{\cal A}_\sigma)\rightarrow  ({\cal A}_\tau+\partial_\tau\lambda,{\cal A}_\sigma+\partial_\sigma\lambda)$, for arbitrary functions $\lambda$ of $\tau$ and $\sigma$, should induce noncommutative gauge transformations on  $A_\mu$:  $A_\mu[{\cal A}_\tau,{\cal A}_\sigma,\phi]\rightarrow A_\mu[{\cal A}_\tau+\partial_\tau\lambda,{\cal A}_\sigma+\partial_\sigma\lambda,\phi]$. For infinitesimal gauge transformations, the latter are given by (\ref{gvrtnsAmu}), with $\Lambda$  a function of $\lambda$,  along with commutative potentials and  their derivatives,  $\Lambda=\Lambda[\lambda,{\cal A}_\tau,{\cal A}_\sigma]$.

The  Seiberg-Witten map can be obtained order by  order in an  expansion in $\theta$,
\beqa  A_\mu&=&  A^{(0)}_\mu+\theta A^{(1)}_\mu+{\cal O}(\theta^2)\cr&&\cr
 \Lambda&=&\Lambda^{(0)}+\theta \Lambda^{(1)}+{\cal O}(\theta^2)\label{axpnnbta}\eeqa
Since we wish to expand the action  $ S_c$, and hence also $y^\mu$, up to second order in $\theta$,  we need  to obtain the  Seiberg-Witten map for $A_\mu$ up to first order.   Except for the inclusion of the scalar field,
the zeroth order expression for the map is uniquely determined from the zeroth order terms in (\ref{oneonesix}).    At lowest  order in $\theta$, $ \Lambda^{(0)}=\lambda$ while the  contributions to  $A^{(0)}_\mu$ from  the commutative gauge potentials are along the tangent directions to the surface, i.e.,
$ A^{(0)}_\mu= h(\tau)\Bigl({\cal A}_\tau\partial_\sigma x_\mu-{\cal A}_\sigma\partial_\tau x_\mu\Bigr)\;+$  the scalar field contribution. 
The scalar field must then be associated with perturbations  normal to the surface; i.e. it's contribution to  $A^{(0)}_\mu$ is proportional to $\phi\,n_\mu$, $\;n_\mu=\Bigl( -a(\tau)a'(\tau),x^1,x^2\Bigr)$.  Thus at zeroth order we may write
\beqa   A^{(0)0}&=& h(\tau)\Bigl( -{\cal A}_\sigma+  {a'(\tau)}a(\tau)\phi\Bigr)\cr&&\cr A^{(0)}_\pm &=&h(\tau) e^{\pm i\sigma}\Bigl(\pm ia(\tau){\cal A}_\tau- a'(\tau){\cal A}_\sigma+a(\tau)\phi\Bigr) \cr&&\cr \Lambda^{(0)}&=&\lambda\label{swzero}\;
\eeqa
To obtain the first order result we demand consistency with (\ref{oneonesix}).  This gives
\beqa   A^{(1)0}&=& h(\tau)\biggl(\frac 1 2 \partial_\tau \Bigl( h(\tau){\cal A}_\sigma^2\Bigr)+a'(\tau)h(\tau) {\cal A}_\tau\partial_\sigma\Bigl(a(\tau)\phi\Bigr)-{\cal A}_\sigma\partial_\tau\Bigr( a'(\tau)h(\tau)a(\tau)\phi\Bigl)\biggr)
 \cr&&\cr A^{(1)}_\pm &=&h(\tau) e^{\pm i\sigma}\biggl(  \mp i \partial_\tau\Bigr( a(\tau)h(\tau) {\cal A}_\tau\Bigr){\cal A}_\sigma\mp i a(\tau)h(\tau) {\cal A}_\tau {\cal F}_{\tau\sigma}\pm i h(\tau) a(\tau) {\cal A}_\tau\phi\cr&&\cr &&\quad+ h(\tau) {\cal A}_\tau \partial_\sigma \Bigl(a(\tau)\phi\Bigr)-{\cal A}_\sigma \partial_\tau\Bigl(h(\tau)a(\tau)\phi\Bigr)+\frac 12 \partial_\tau\Bigr( a'(\tau)h(\tau) {\cal A}_\sigma^2 \Bigr)-\frac12 a(\tau)h(\tau){\cal A}_\tau^2\biggr) 
\cr&&\cr \Lambda^{(1)}&=&- h(\tau) {\cal A}_\sigma\partial_\tau\lambda\label{swone}
\eeqa
where ${\cal F}_{\tau\sigma}=\partial_\tau {\cal A}_\sigma-\partial_\sigma {\cal A}_\tau$ is the $U(1)$ gauge field on the surface.

To obtain the  lowest order action action for the  scalar field and gauge field on the two-dimensional space-time manifold, we  substitute (\ref{yxbta}) and  (\ref{axpnnbta})-(\ref{swone}) into (\ref{cmtvlmtscnqt}) and keep only up to quadratic terms in the perturbation parameter $\theta$.  After some work, we  get 
\beqa
 S_c(y)&=&\frac {\theta^2}{g_c^2}\int d\tau d\sigma\sqrt{-{\tt g}}\,\Bigl(\frac 14 {\cal F}^{\tt a b}{\cal F}_{\tt ab}-\frac 12\partial^{\tt a}\phi\partial_{\tt a}\phi- \frac 1{2}m^2\phi^2 \Bigr)  \;+\; S_c(x)\;,\label{scgnrlfrm}\eeqa
where ${\tt g}$ is again the determinant of the induced metric  ${\tt g}_{\tt ab}$ and $m$ denotes a background-dependent mass for the scalar field. This  is the usual expression for the action of a  scalar field and gauge field, except for the sign in front of the electric field contribution.  However, the electric field, which  is   nondynamical in two dimensions, is decoupled from the scalar and therefore of no concern for dynamics.  The explicit expression for the action in terms of the scale factor $a(\tau)=\cos\tau$ is
\beqa
S_c(y)&=&\frac {\theta^2}{g_c^2}\int d\tau d\sigma\,\Bigl(-\frac 1{2a(\tau)^2}{\cal F}_{\tau\sigma}^2+\frac 12(\partial_\tau\phi)^2-\frac 12(\partial_\sigma\phi)^2 + \frac 1{a(\tau)^2}\phi^2 \Bigr)  \;+\; S_c(x)\;,\label{1.19}
\eeqa
where the action evaluated for the classical solution $y^\mu=x^\mu$ is $S_c(x)=-\frac {1}{2g_c^2}\int d\tau d\sigma\cos^2\tau=-\frac {\pi^2}{2g_c^2}$.    From (\ref{1.19}), the scalar field is tachyonic. 
The system is thus unstable with respect to perturbations normal to the surface.
 The tachyonic mass-squared $m^2$ is scale-dependent.  By comparing (\ref{scgnrlfrm}) to (\ref{1.19}), one gets that
\be m^2=-\frac 2{a(\tau)^4}\;\label{twotwo8}\ee
The tachyonic mass is inversely proportional to the scale-squared and is singular in the limit  $\tau$ tends to $-\frac \pi 2$ and  $ \frac \pi 2$.

\section{Adding a Cubic term}
\setcounter{equation}{0}
\subsection{Modified matrix equations and two well known solutions}

More solutions to the matrix model are possible upon  including a cubic term in the action, which introduces a free parameter to the theory.   With this in mind, we replace (\ref{mmactnnqt}) by 
\be S(Y)=\frac 1{g^2}{\rm Tr}\Bigl(-\frac 14 [Y_\mu, Y_\nu] [Y^\mu,Y^\nu] +\frac 23 i \alpha \epsilon_{\mu\nu\lambda}Y^\mu Y^\nu Y^\lambda\Bigr)\;,\label{mmact}\ee where $\alpha$ is the free parameter.  Our convention for the Levi-Cevita tensor is $\epsilon_{012}=1$.  The equations of motion  now  read
\be [ [Y_\mu,Y_\nu],Y^\nu]+i\alpha \epsilon_{\mu\nu\lambda}[Y^\nu,Y^\lambda] =0\label{eqofmotnctrm} \;\ee They preserve the symmetries i-iii) of (\ref{eqofmotnqt}).
 
There are two well known solutions to these equations, and they are associated with finite dimensional Lie algebras.  One  is  the noncommutative   de Sitter solution.\cite{Ho:2000fy},\cite{Jurman:2013ota}  For this one sets $ Y^\mu=X^\mu$, where $X^\mu$ are the generators   of
 the $2+1$ Lorentz group
 \be  [ X^\mu, X^\nu]= i \alpha \epsilon^{\mu\nu\lambda} X_{\lambda}  \;\label{ncds}\ee    An irreducible representation results upon setting  the  Casimir of the algebra $X^\mu X_{\mu}$ equal to a constant times the identity. This solution is the Lorentzian space analogue of the fuzzy sphere.\cite{Madore:1991bw}-\cite{Iso:2001mg}

Another solution  is the noncommutative cylinder.\cite{Chaichian:2000ia},\cite{Balachandran:2004yh},\cite{me}  It is given by $Y^\mu=X^\mu$, where $X^\mu$ now generate  the two dimensional Euclidean group,
 \be  [X_{0},X_{\pm}]=\pm  2 \alpha X_{\pm}  \qquad\qquad   [X_{+},X_{-}]=0\;, \label{ncbeta0}\ee with $X_{\pm}=X_{1} \pm i X_{2}$.   The algebra possesses two central elements $X_{+}X_{-}$  and $\;\exp{\Bigl(\frac {\pi i}\alpha X_{0}\Bigr) }$, whose eigenvalues determine the   irreducible  representations.  The eigenvalue of $X_{+}X_{-}$  is the radius  of the noncommutative cylinder, while the eigenvalues of the `time' operator $X_0$ are regularly spaced.

We  argue that there can be a third solution to (\ref{eqofmotnctrm}), which is just a deformation of the previously proposed solution  to (\ref{eqofmotnqt}).  For this we  again examine the commutative limit. 

\subsection{Solutions in the commutative limit}

We again introduce the noncommutativity parameter $\theta$, with    the commutative limit corresponding to  $\theta\rightarrow 0 $.  In order that the both terms in  (\ref{eqofmotnctrm}) survive in the limit, we need that  $\alpha$ goes to zero  and is  of order $\theta$  as $\theta\rightarrow 0$:  \be \alpha\rightarrow \upsilon \theta\;, \qquad \upsilon\;\;{\rm finite}\ee  Then (\ref{eqofmotnctrm}) becomes
\be \{\{y_\mu,y_\nu\},y^\nu\}+\upsilon \epsilon_{\mu\nu\rho}\{y^\nu,y^\rho\} =0 \label{clmeas}
\;,\ee  which generalizes (\ref{clmeasnqt}).  

We denote solutions to (\ref{clmeas}) by  $y^\mu=x^\mu$. The commutative analogues of (\ref{ncds}) and (\ref{ncbeta0}) are examples of solutions, and they can be expressed in terms of the functions $a$ and $h$ appearing  in (\ref{cmtvcnstrnt}) and (\ref{gnrlpbs}).   The commutative limit of  (\ref{ncds})  is   \be a^2(x^0)=\frac 1{\upsilon^2}+(x^0)^2\qquad\quad h(x^0)=\upsilon\;,\label{clds2soln}\ee
while the  commutative limit of  (\ref{ncbeta0}) is  \be a=\frac 1{2\upsilon}\qquad\quad h=2\upsilon\label{cyldrsln}\ee The solutions (\ref{clds2soln}) and  (\ref{cyldrsln}) represent the $2D$ de Sitter universe and static  universe, respectively.
The $\upsilon $  dependence was inserted  in $a(x^0)$ in  (\ref{clds2soln}) and  (\ref{cyldrsln}) in order  that 
 the condition (\ref{his1vrsmt}) is satisfied, however this is not a necessary condition to solve   (\ref{clmeas}).     Both solutions  are singular in the limit $\upsilon\rightarrow 0$.  Also in both cases they lead to linear Poisson brackets. 
 
 More generally, one has a solution to  (\ref{clmeas}) if the two functions $a$ and $h$ in (\ref{cmtvcnstrnt}) and (\ref{gnrlpbs}) satisfy
\be  \Bigl ((aa'h)'+h-2\upsilon\Bigr)\,h=0\qquad \quad  \Bigl(2ha'+ah'-2\upsilon a'\Bigr)\,ah=0\label{3difeqspct}\,\ee 
These equations generalize (\ref{3difeqsnqt}).   They are satisfied  for (\ref{clds2soln}) and for (\ref{cyldrsln}).
In addition to them, one can obtain  solutions to (\ref{3difeqspct}) which  are   deformations of  (\ref{afhnqt}).  Upon imposing the condition (\ref{his1vrsmt}), we now get the following equation for the scale factor 
\be \frac {a''}a= \Bigl(\frac{a'}a\Bigr)^2
-\frac 1{a^2}+\frac{ 2\upsilon} a (1-a'^2)^{\frac 32}\label{scndrdrfra}
\ee   This yields the integral of the motion $a/{\sqrt{1-a'^2}}-\upsilon a^2$, and as was the case with $\upsilon=0$, it can be associated with the energy of a bosonic string, as we shall see later. This yields the integral of the motion leads to  the following Friedmann-type equation for the scale factor,
\be \Bigl(\frac{a'}a\Bigr)^2-\frac 1{a^2}= -\frac 1{({\cal E}+\upsilon a^2)^2}\;,\label{frdmneq}\ee  ${\cal E}$ being the integration constant.\footnote{  In comparing with the usual expression for cosmological evolution (in four space-time dimensions), the
right hand side of (\ref{frdmneq})  behaves like a negative scale-dependent energy density.  The latter  is most significant at small scales.   Moreover, from (\ref{scndrdrfra}) one can also identify a scale-dependent pressure term.  It too can be negative, thus mimicking dark energy.  Of course, it would be more appropriate to compare the results with cosmological solutions to Einstein gravity in two space-time dimensions.  However,   Einstein gravity does not exist in  two space-time dimensions;  the Einstein tensor identically vanishes, meaning that the theory cannot support a non vanishing energy-momentum source (except for a cosmological term).  On the other hand, interpretations may be possible in the context of alternative formulations of gravity in  two space-time dimensions,\cite{Jackiw:1995hb},\cite{Grumiller:2002nm} including an interesting $\epsilon\to 0$ limit of Einstein gravity in $2+\epsilon$ dimensions.\cite{Grumiller:2007wb}}  
The solutions to  (\ref{frdmneq}) resemble familiar cosmological space-times.
 Solutions   can be
expressed in terms of  inverse elliptic integrals.  For the boundary condition, let us assume that $a$ has a turning point at $x^0=0$.
 Then the resulting solutions can  describe closed, stationary or open space-times, the choice depending on the value of $\upsilon$. Closed two-dimensional space-times, having initial and final singularities at some $ x^0=\pm \tau_0$, occur for $\upsilon<\frac 12$ (including negative $\upsilon$).   An example, discussed in the previous section, is the case of  $\upsilon= 0$, whose solution  is given by the simple expression (\ref{afhnqt}).  It, as well as two other examples of solutions for $\upsilon<\frac 12$, are exhibited in figure 1.    The case of $\upsilon=\frac 12$ coincides with  the static or cylindrical space-time solution   (\ref{cyldrsln}), and is shown in figure 2(a).   Open universe solutions are recovered for $\upsilon>\frac 12$, examples of which are shown in figures 2(b) and 2(c).   The   case $\upsilon= 1$ coincides with  the de Sitter solution   (\ref{clds2soln}), shown in figure 2(c).   There are simple expressions for the solutions when $\upsilon= 0,\frac 12$ and $1$, pictured in figures 1(b), 2(a) and 2(c).
\begin{figure}
\centering
\begin{subfigure}{.325\textwidth}
  \centering
  \includegraphics[height=1.3in,width=1.8in,angle=0]{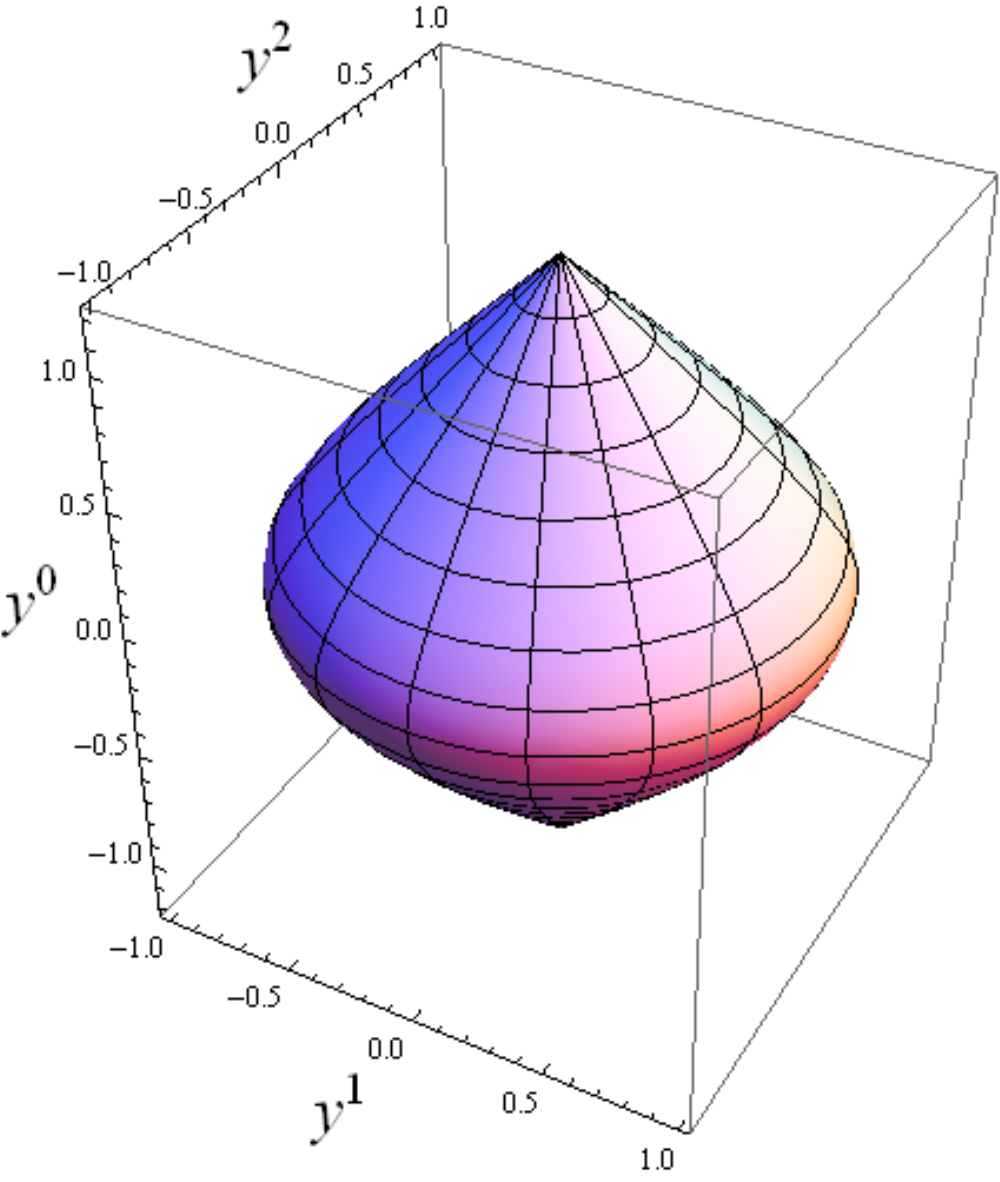}
  \caption{$\upsilon=-1.5$}
  \label{fig:sub1}
\end{subfigure}%
\begin{subfigure}{.325\textwidth}
  \centering
  \includegraphics[height=1.8in,width=1.75in,angle=0]{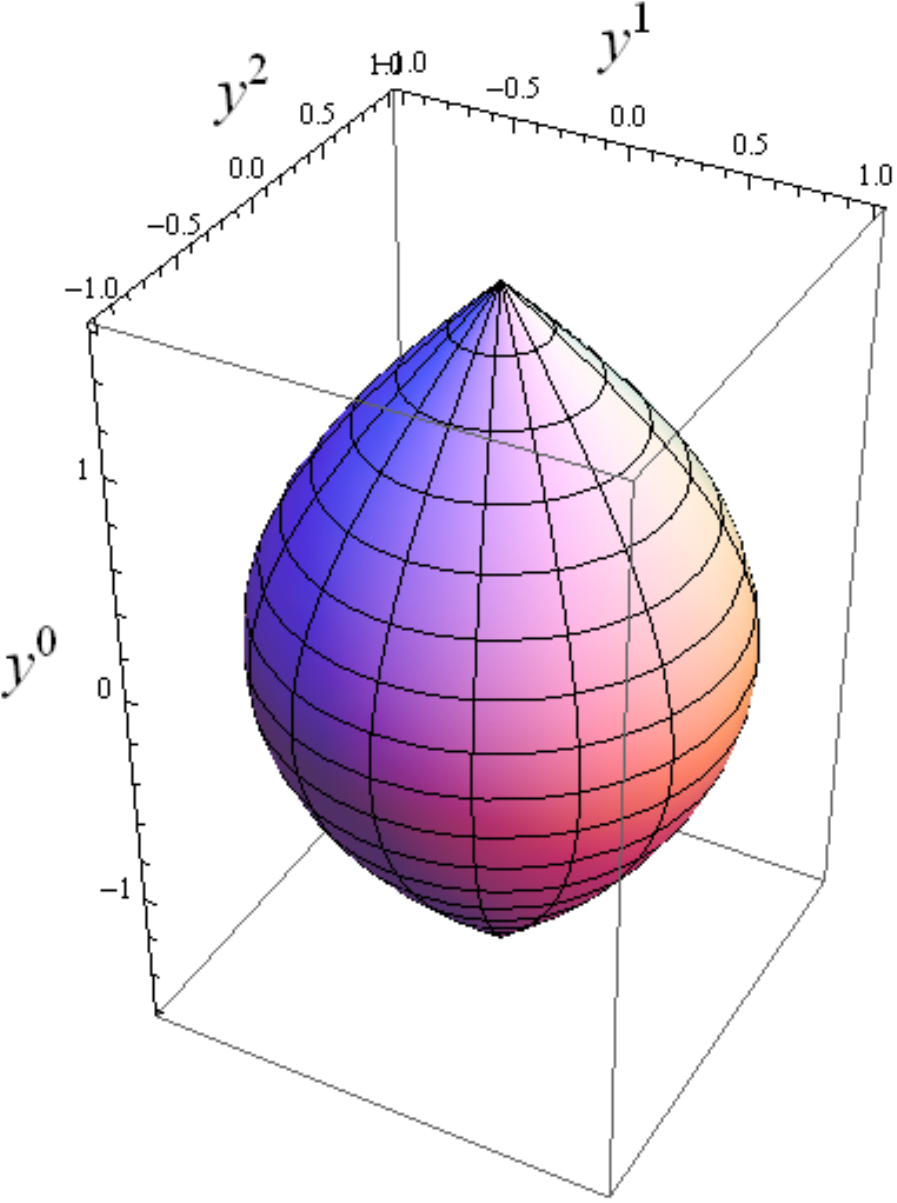}
  \caption{$\upsilon=0$}
  \label{fig:sub1}
\end{subfigure}%
\begin{subfigure}{.325\textwidth}
  \centering
  \includegraphics[height=2.4in,width=1.8in]{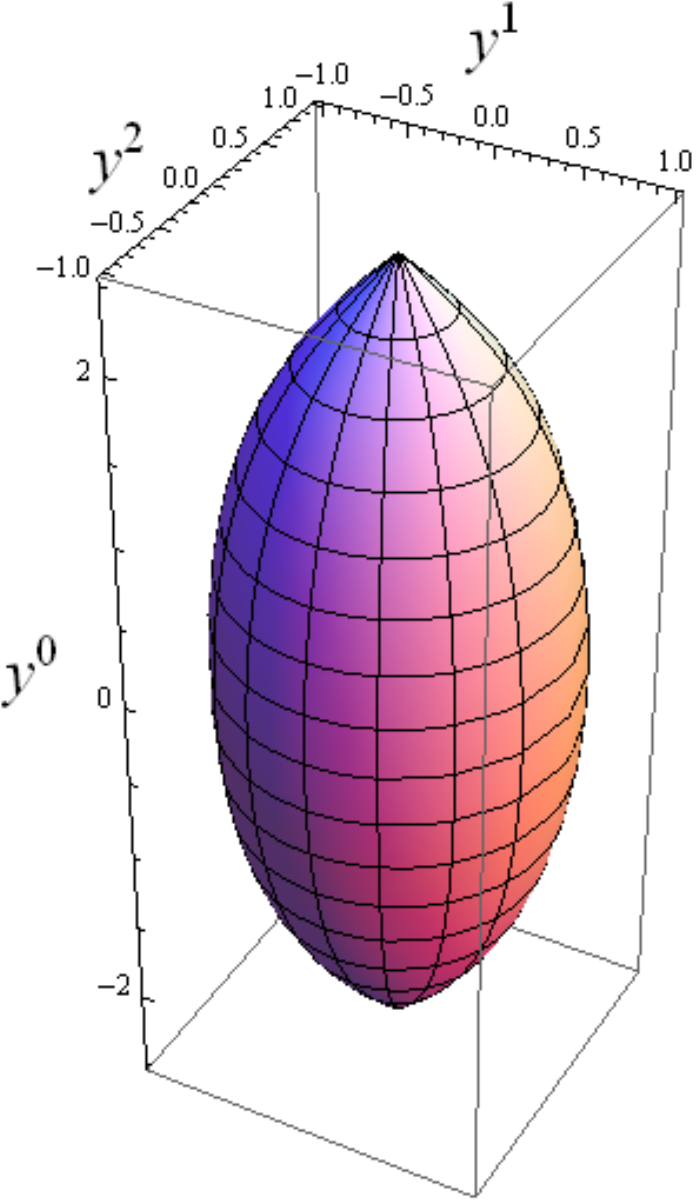}
  \caption{$\upsilon=.4$}
  \label{fig:sub2}
\end{subfigure}
\caption {Parametric plots of the closed universe solutions in the three-dimensional embedding space (with time along the vertical direction) for three different values of  $ \upsilon <\frac 12$.  The boundary conditions are $a(0)=1$ and $a'(0)=0$.}
\label{fig:test}
\end{figure}
\begin{figure}
\centering
\begin{subfigure}{.325\textwidth}
  \centering
  \includegraphics[height=1.7in,width=1in,angle=0]{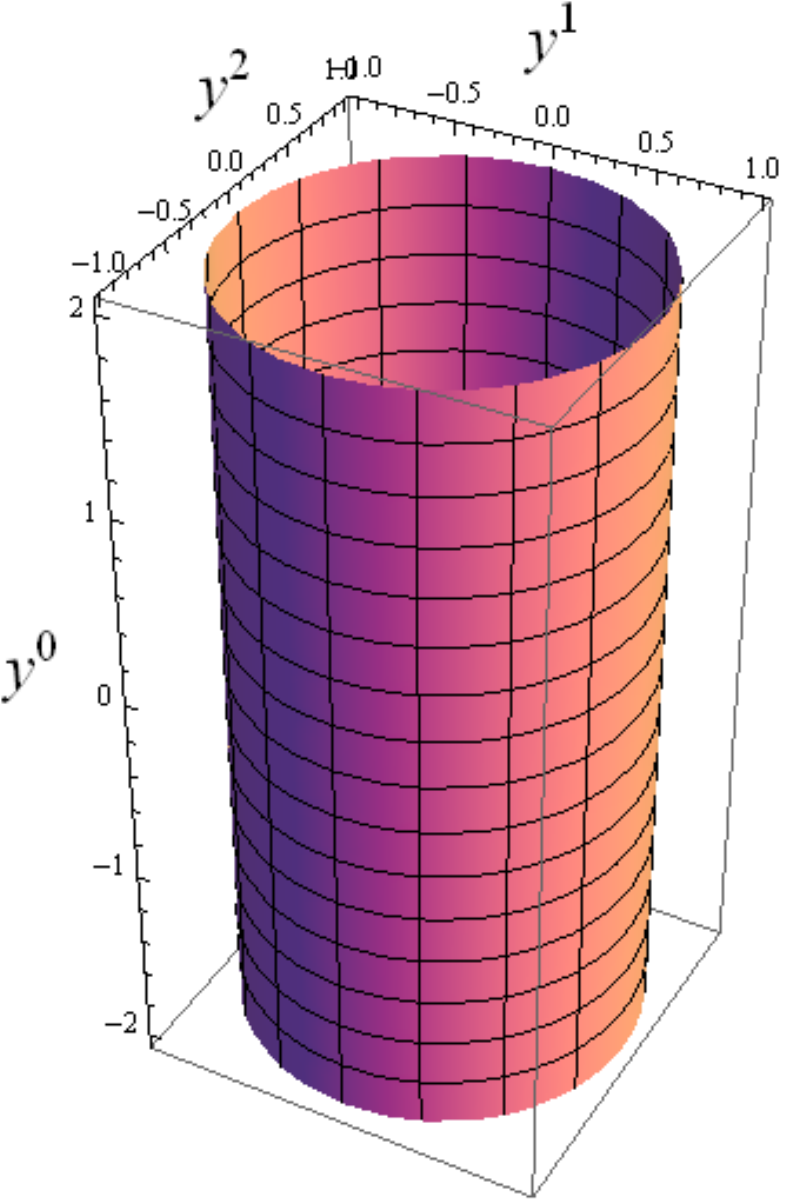}
  \caption{$\upsilon=.5$}
  \label{fig:sub1}
\end{subfigure}%
\begin{subfigure}{.325\textwidth}
  \centering
  \includegraphics[height=1.9in,width=1.5in,angle=0]{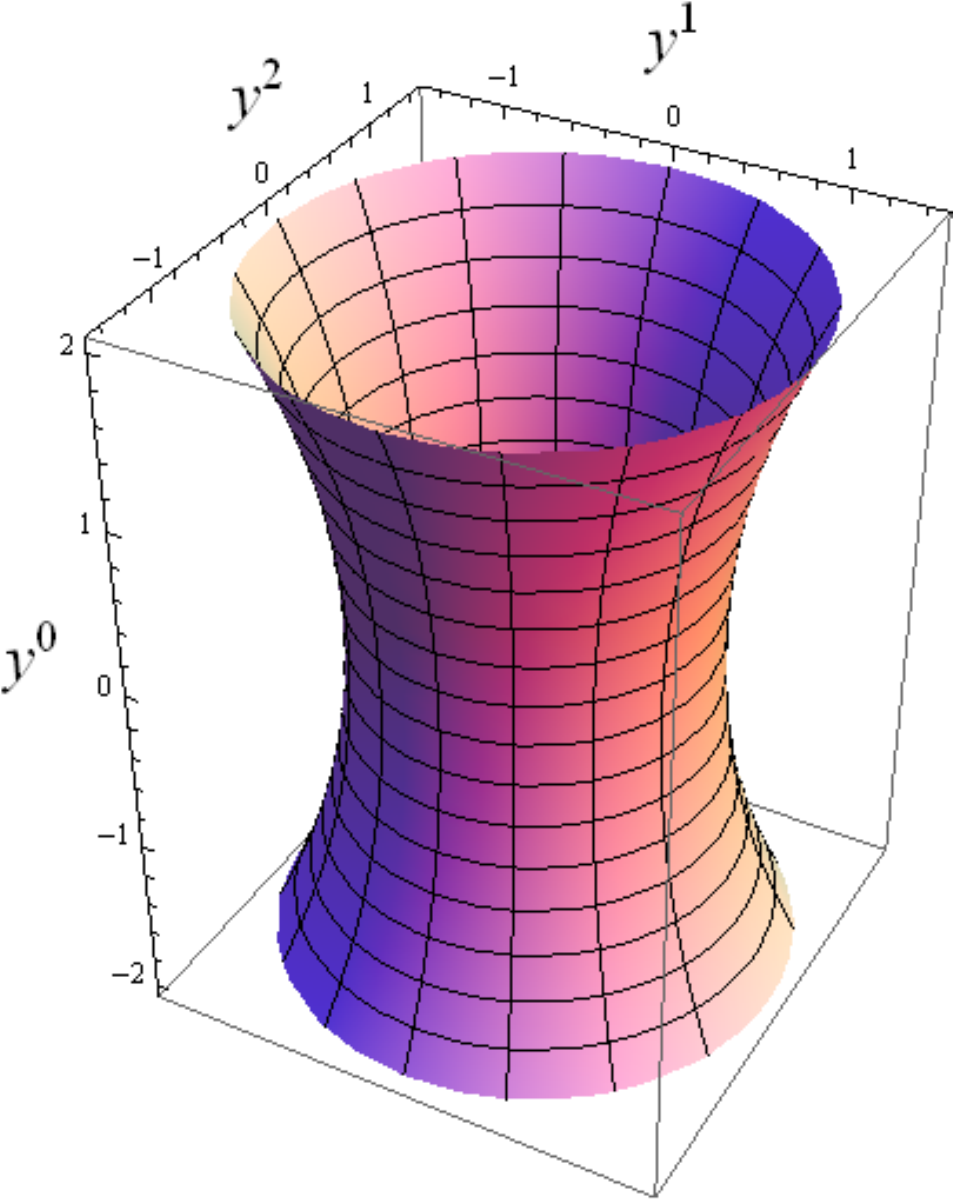}
  \caption{$\upsilon=.6$}
  \label{fig:sub1}
\end{subfigure}%
\begin{subfigure}{.325\textwidth}
  \centering
  \includegraphics[height=2.2in,width=2.2in]{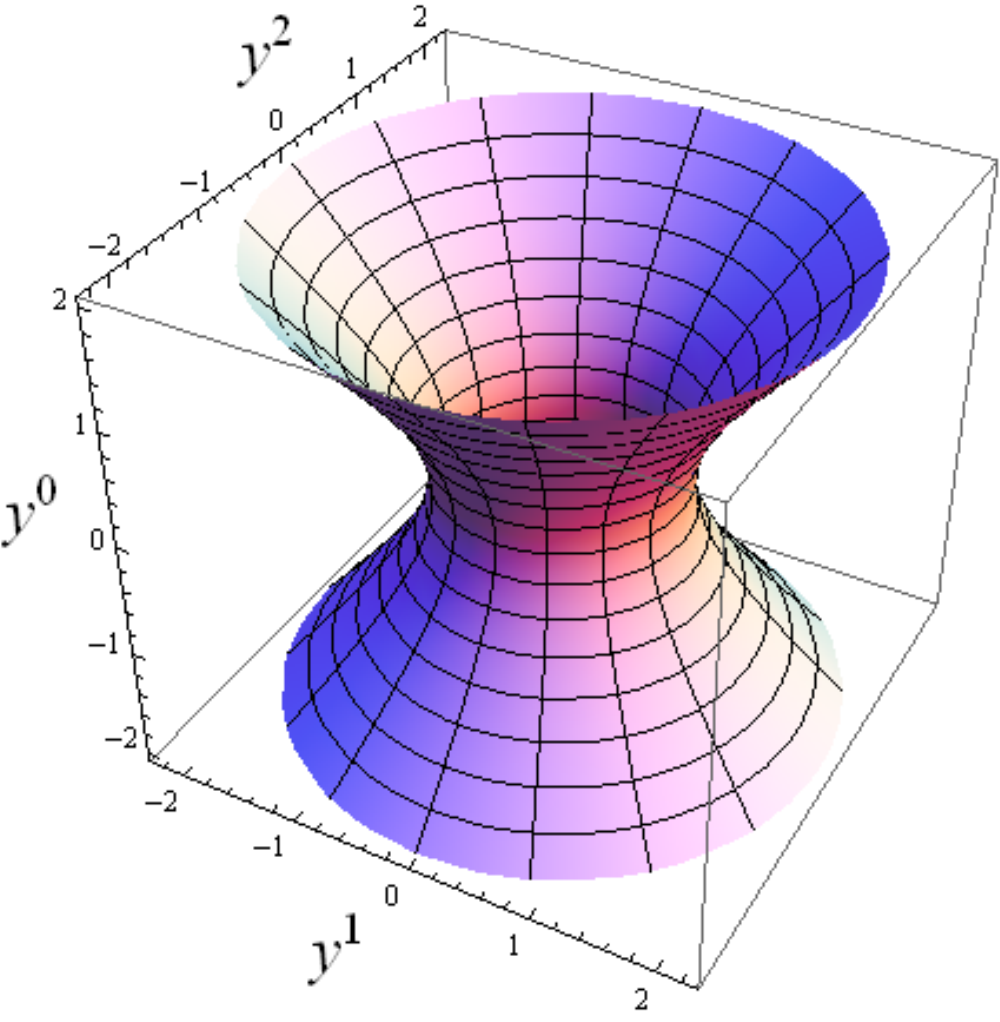}
  \caption{$\upsilon=1$}
  \label{fig:sub2}
\end{subfigure}
\caption {Parametric plots of the static universe  solution $\upsilon =\frac 12$ and open universe solutions for two different values of  $\upsilon >\frac 12$ in the three-dimensional embedding space  (with time along the vertical direction).   The boundary conditions are $a(0)=1$ and $a'(0)=0$.}
\label{fig:test}
\end{figure}

In summary,  we have found solutions to (\ref{clmeas}) of the form (\ref{cmtvcnstrnt}) and (\ref{gnrlpbs}).  In parametric form they were given by (\ref{ncprtztn}) and  (\ref{pbtauphi}). For generic values  of $\upsilon$, there are three distinct solutions.  They are   the de Sitter universe (\ref{clds2soln}), the static universe (\ref{cyldrsln}) and deformations of (\ref{afhnqt}), which are solved by inverse elliptic integrals.  At some special values of $\upsilon$, there are fewer than three distinct solutions. For $\upsilon= 0$ there is only one solution, (\ref{afhnqt}), corresponding to a closed universe.  For  $\upsilon=\frac 12$ and $1$, there are two solutions,   (\ref{clds2soln}) and  (\ref{cyldrsln}),  corresponding to the static universe  and de Sitter universe, respectively.
With  the exceptions of   $\upsilon=\frac 12$ and $1$,  the  solutions yield a nonlinear Poisson  bracket algebra (\ref{gnrlpbs}).  So except for these two cases, the corresponding matrix solutions are  nontrivial.  They are not  investigated here. 

We address the question of stability of these solutions first  from the perspective of  classical strings  and then from the perspective of the matrix models.

\subsection{Classical string perspective} 
 As with the case of $\upsilon=0$, the parametric expression (\ref{ncprtztn}) solves the equations of motion for a classical closed bosonic string, in addition to solving (\ref{clmeas}).  However when $\upsilon\ne0$,   we must add a term, which we denote by $S_{NS}$, to   the standard  Nambu-Goto action:
\be S_{string}=S_{NG} +S_{NS}\;,\qquad\quad S_{NS}=-\frac{\upsilon {\cal  T}}{3}\int  \epsilon_{\mu\nu\rho}y^\mu dy^\nu\wedge dy^\rho\label{clnbns}\ee  It can be regarded as a coupling to a Neveu-Schwarz field of the from $B_{\mu\nu}\propto \epsilon_{\mu\nu\lambda}y^\lambda$. Both terms in the action (\ref{clstactn}) are reparametrization invariant, and respect the Poincar\'e symmetry in $2+1$  space-time.
The Nambu string equations (\ref{cleomnqt})  are now modified to
\be  \Delta y_\mu + 2\upsilon n_\mu=0\;,\label{cleom}\ee where  $n_\mu=\frac 1{2{\sqrt{-{\tt g}}}}\epsilon^{\tt ab}\epsilon_{\mu\nu\rho}\partial _{\tt a} y^\nu\partial_{\tt b} y^\rho$ is a space-like unit vector normal to the world sheet and $\epsilon^{\tau\sigma}=-\epsilon^{\sigma\tau}=1$.  The string equations (\ref{cleom}) are identical to the equations  (\ref{clmeas}) when  the Poisson structure on the world sheet involves the metric tensor according to (\ref{rltpbmtrc}). 

Once again, the string  equations of motion imply the existence of a conserved current  on the world sheet.  In comparing with  (\ref{panu}), it has an additional term,
\be  p^{\tt a}_\mu=-{\cal T}\sqrt{-{\tt g}} {\tt g}^{\tt ab}\partial_{\tt b }y_\mu+\upsilon{\cal T}\epsilon^{\tt ab}\epsilon_{\mu\nu\rho}\, y^\nu\partial_{\tt b} y^\rho\label{panunew}
\ee  Then the string energy   (\ref{strngnrg}) evaluated for any solution $y^\mu=x^\mu$ of the form  (\ref{ncprtztn}), also aquires an extra term 
\be E=2\pi {\cal T}a(z^0)\Bigl(\frac 1{\sqrt{1-a'(z^0)^2}}-\upsilon a(z^0)\Bigr) \ee
It is proportional to the integration constant ${\cal E}$ appearing in (\ref{frdmneq}), $E=2\pi {\cal TE}$. From the  choice of boundary conditions used in figures 1 and 2, ${\cal E}=1-\upsilon$.

Let us compare the string energies of the three different types of solutions.
The string energy vanishes when evaluated for the de Sitter solution  (\ref{clds2soln}) is zero, $E|_{dS ^2}=0$, and so this solution is  degenerate with the vacuum.  For the case of the cylindrical space-time solution it is instead proportional to $\upsilon$,  $E|_{R\times S}=2\pi {\cal T}\upsilon$.  The remaining family of solutions are  associated with   inverse elliptic integrals, and describe  closed universes for $\upsilon<\frac 12$, the static universe for $\upsilon=\frac 12$ and open universes for $\upsilon>\frac 12$.  Let  us again adopt    the boundary conditions  used in figures 1 and 2, i.e., $a(0)=1$ and $a'(0)=0$. The energy for this family of solutions is   $E|_{{\rm elliptic}^{-1}}=2\pi {\cal T}(1-\upsilon)$.  We plot the string energies associated with the three different types of solutions in figure 3.  The cylindrical space-time solution is the lowest energy configuration in the region $\upsilon<0$.   On the other hand, the $dS^2$  solution has the minimum energy  for $0<\upsilon\le1$, and it is degenerate with the vacuum solution.  The family of open universe solutions is the minimum energy configuration for $\upsilon>1$.  The closed universe solution never has the least energy solution, except for the case $\upsilon=0$, when it  is the only nontrivial of the three types of  solutions to survive. However it is unstable with respect to decay to  the vacuum.
\begin{figure}[placement h]
\begin{center}
\includegraphics[height=2in,width=2.5in,angle=0]{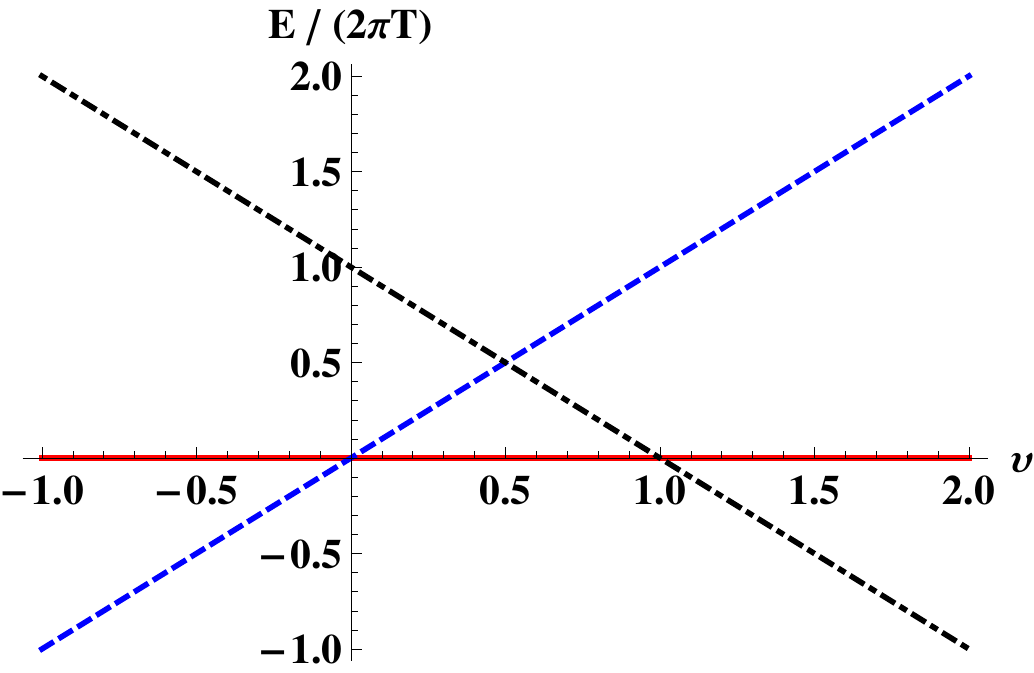}
\caption {Plots of the string energy (divided by $2\pi{\cal T}$) for the three types of solutions as a function of $\upsilon$.  The energy for the de Sitter solution is zero  (red solid line).  The energy for the cylindrical solution is given by the blue dashed line. (The de Sitter solution and cylinder solution are singular at   $\upsilon=0$). The  energy of the family of solutions given by   inverse elliptic integrals is given by the black dashed-dotted  line. It corresponds to  closed universes for $\upsilon<\frac 12$, the static universe for $\upsilon=\frac 12$ and open universes for $\upsilon>\frac 12$.  The   boundary conditions are $a(0)=1$ and $a'(0)=0$.  } 
\end{center}

\end{figure}

\subsection{Matrix model perspective}

 Instead of the string action (\ref{clnbns}), we now consider the commutative limit of the matrix action (\ref{mmact}):
\be
 S^{{\rm total}}_c(y)= S_c(y)+S^{(3)}_c(y) \;,\qquad\quad S^{(3)}_c(y) = -\frac {\upsilon }{3 g_c^2}\int d\mu(\tau,\phi)\, \epsilon_{\mu\nu\lambda}\,y^\mu\{y^\nu, y^\lambda\}\;,\label{cmtvlmtsc}
\ee  where $ S_c(y)$ was given in (\ref{cmtvlmtscnqt}) and $ d\mu(\tau,\sigma)$ is again an invariant  integration measure on the world sheet.  Its  resulting  equations of motion are  (\ref{clmeas}), and now we don't have to impose    (\ref{his1vrsmt}) for this purpose. 
We wish to evaluate this action  for small perturbations  about the three types of solutions  to (\ref{3difeqspct}). The perturbations can again be expressed in terms of noncommutative potentials $A_\mu$, as in (\ref{yxbta}), which can then be rewritten as functions of commutative gauge potentials,  $({\cal A}_\tau,{\cal A}_\sigma)$ and a scalar field $\phi$ on  the surface using  the Seiberg-Witten map  (\ref{axpnnbta})-(\ref{swone}).  Upon substituting the map into (\ref{cmtvlmtsc}), we get the action for small fluctuations in terms of $({\cal A}_\tau,{\cal A}_\sigma)$ and  $\phi$.  For all three types of solutions, we recover the terms appearing in  (\ref{scgnrlfrm}) describing the  scalar field and  electromagnetism on the two-dimensional space-time manifold (with $m^2=0$ for the cylindrical solutions). As before the kinetic energy terms for the gauge and scalar field appear with opposite sign. Now when $\upsilon\ne 0$ we obtain an additional coupling term,
\beqa
 S_c(y)&=&\frac {\theta^2}{g_c^2}\int d\tau d\sigma\sqrt{-{\tt g}}\,\Bigl(\frac 14 {\cal F}^{\tt a b}{\cal F}_{\tt ab}-\frac 12\partial^{\tt a}\phi\partial_{\tt a}\phi- \frac 1{2}m^2\phi^2 +\frac{2\upsilon}{\sqrt{-{\tt g}}}\,\phi {\cal F}_{\tau\sigma}\Bigr)  \;+\; S_c(x)\;,\label{scgnrlfrmplscpl}\eeqa
The explicit expression in terms of the scale factor $a(\tau)$ is
\beqa
 S_c(y)&=&\frac {\theta^2}{g_c^2}\int d\tau d\sigma\,\Biggl(-\frac 1{2a(\tau) \sqrt{1-a'(\tau)^2}}{\cal F}_{\tau\sigma}^2+\frac {a(\tau)}{2\sqrt{1-a'(\tau)^2}}(\partial_\tau\phi)^2-\frac {\sqrt{1-a'(\tau)^2}}{2a(\tau) }(\partial_\sigma\phi)^2\cr &&\cr&& \qquad\qquad +\biggl( \frac1 {a(\tau)\sqrt{1-a'(\tau)^2}}-2\upsilon\biggl) \phi^2+2\upsilon\phi {\cal F}_{\tau\sigma}\Biggr)  \;+\; S_c(x)\;,\label{gnlcylactn}
\eeqa
where 
the action evaluated for any of the three classical solutions $y^\mu=x^\mu$ can be written as
 \be S_c(x)= -\frac {\pi}{  g_c^2}\int d\tau\,a(\tau)\Bigl(4\tau \upsilon a'(\tau)+\sqrt{1-a'(\tau)^2}\Bigr)\ee
This quantity is divergent for the  solutions  describing open and cylindrical space-times. 
The results agree with (\ref{1.19}) in the limit $\upsilon\rightarrow 0$.  Upon comparing (\ref{scgnrlfrmplscpl}) and (\ref{gnlcylactn}), the mass-squared  of the scalar field is
\be m^2=-\frac{2}{a(\tau)\sqrt{1-a'(\tau)^2}}\biggl( \frac1 {a(\tau)\sqrt{1-a'(\tau)^2}}-2\upsilon\biggl)\label{gnrlsms}\ee  Evaluating it for the three types of solutions, one finds that it vanishes for the case of cylindrical space-time solutions (\ref{cyldrsln}),  it has a positive value of   $2\upsilon^2$ for  the case of de Sitter
 solutions (\ref{clds2soln}), while its value  {\it  and sign} are scale-dependent
for the case of general solutions to (\ref{frdmneq}) expressed in terms of inverse elliptic integrals.  In the latter case,  $m^2=2\biggl(\upsilon^2-\frac{{\cal E}^2}{a(\tau)^4}\biggr)$, where ${\cal E}$ is again the integration constant appearing in (\ref{frdmneq}).

 One additional step is needed  to do the stability analysis due to the  coupling of the scalar field  to the nondynamical gauge field, which is present when $\upsilon\ne 0$.  The gauge field can be eliminated using its equation of motion, $\frac{{\cal F}_{\tau\sigma}}{a(\tau)\sqrt{1-a'(\tau)^2}}=2\upsilon\phi+{\rm constant}$.
After substituting back into the action, the  mass-squared for the scalar field (\ref{gnrlsms}) gets modified to
\be m^2_{eff}=-2\Biggl\{\upsilon^2+\biggl(\frac{1} {a(\tau)\sqrt{1-a'(\tau)^2}}-\upsilon\biggr)^2
\Biggr\}
\ee
{\it Thus the scalar field is  tachyonic for all three types of solutions.}  Moreover, the effective mass-squared is scale-dependent for the family of solutions given in terms of inverse elliptic integrals.  The results for the three cases are:
\be m^2_{eff}=\left\{ \matrix{-4\upsilon^2\;,& {\rm cylindrical\; solution}\cr
-2\upsilon^2 \;,&dS^2\;{\rm solution}\cr -2\Bigl(\upsilon^2+\frac{{\cal E}^2}{a(\tau)^4}\Bigr)\;,&{\rm  inverse\; elliptic\; integral\; solution}
\cr
}\right. \label{uncmas}\ee
Recall that the cylindrical solution and $dS^2$ solution are  singular in the  limit $\upsilon\rightarrow 0$, so the scalar field can never be massless.
The result for the inverse elliptic integral solution agrees with what we found previously (\ref{twotwo8}) in the absence of the cubic term.  
Thus all of the solutions examined here were found to be unstable with respect to perturbations  normal to the surface.  

\section{Concluding remarks}
The results obtained here indicate that instabilities may be a common feature of the cosmological solutions to simple matrix models. 
Additional terms must be included in the IKKT matrix model action in order to cure the above instabilities. The quartic term
${\rm Tr}(Y_\mu Y_\nu)^2$, with a suitably adjusted coefficient, was shown to stabilize the cylindrical space-time solution.\cite{me}
The quadratic term ${\rm Tr}(Y_\mu Y_\nu)$ is sufficient to cure the instability of the  noncommutative $dS^2$ solution.\cite{LeiAndrea} (A quadratic term was also included in  \cite{Freedman:2004xg} and it played a role of a cosmological term in the Friedmann equations.)  Similar such stabilization terms should be possible for the inverse elliptic integral solutions associated with closed and open universes.
The tachyonic   mass was found to be scale-dependent  for the  inverse elliptic integral solutions, and is larger at  smaller distance  scales. The  inclusion of extra terms in this case could produce a transition to a stable solution at a certain scale.  This may be of use for inflationary models, as it is analogous to the transition from an inflationary to non-inflationary phase. 

Of course, it is  of interest to generalize the $2D$ solutions studied here to higher dimensions, and see whether realistic cosmological space-times arise from such models.  It may be a nontrivial problem, however, to insure full rotation invariance in an arbitrary number of dimensions.  In that case we want to replace (\ref{cmtvcnstrnt}) by $ ( x^1)^2+ ( x^2)^2+\cdot\cdot\cdot + ( x^d)^2=a^2(x^0)$, $d>2$, but the generalization of the Poisson brackets (\ref{gnrlpbs}) to more than two spatial embedding dimensions is not obvious.  On the other hand, one can examine solutions which are obtained by taking products of the lower dimensional noncommutative spaces examined here.
 We plan to explore these issues in coming works.

\bigskip

{\Large {\bf Acknowledgments} }

\noindent
I am very grateful to A. Chaney, D. Grumiller, B. Harms, L. Lu, A. Pinzul and H. Steinacker for valuable discussions.

 \end{document}